\documentclass[11pt,a4paper]{article}
\usepackage{jcappub}
\usepackage[utf8]{inputenc}
\usepackage{url}
\usepackage{hyperref}
\usepackage{natbib}

\title{Cosmology and stellar equilibrium using Newtonian hydrodynamics with general relativistic pressure}

\author[a]{P.~O.~Baqui,}
\author[a]{J.~C.~Fabris,}
\author[a]{and O.~F.~Piattella}

\affiliation[a]{Department of Physics, Universidade Federal do Esp\'irito Santo,\\
avenida Ferrari 514, 29075-910 Vit\'oria, Esp\'irito Santo, Brazil}

\emailAdd{pedrobaqui@gmail.com}
\emailAdd{oliver.piattella@pq.cnpq.br}
\emailAdd{fabris@pq.cnpq.br}

\abstract{We revisit the analysis made by Hwang and Noh [JCAP 1310 (2013)] aiming the construction of a Newtonian set of equations incorporating pressure effects typical of the General Relativity theory. We explicitly derive the Hwang-Noh equations, comparing them with similar computations found in the literature. Then, we investigate $i)$ the cosmological expansion, $ii)$ linear cosmological perturbations theory and $iii)$ stellar equilibrium by using the new set of equations and comparing the results with those coming from the usual Newtonian theory, from the Neo-Newtonian theory and from the General Relativity theory. We show that the predictions for the background evolution of the Universe are deeply changed with respect to the General Relativity theory: the acceleration of the Universe is achieved with positive pressure. On the other hand, the behaviour of small cosmological perturbations reproduces the one found in the relativistic context, even if only at small scales. We argue that this last result may open new possibilities for numerical simulations for structure formation in the Universe. Finally, the properties of neutron stars are qualitatively reproduced by Hwang-Noh equations, but the upper mass limit is at least one order of magnitude higher than the one obtained in General Relativity.}


\arxivnumber{1512.09056}

\begin{document}
\maketitle

\flushbottom


\section{Introduction}

The General Relativity theory (GR) is the modern theory of gravitation that has replaced Newtonian theory of gravity, which in principle was very successful. However, Newtonian gravity is still largely used, especially in the study of astrophysical and cosmological phenomena. In fact, the corrections given by GR theory to Newtonian gravity are typically extremely small, except in the regime of high velocities and strong gravitational fields. GR becomes essential, for example, in the description of very compact objects (like neutron stars and black holes), in the primordial Universe, and in the present phase of accelerated expansion of the Universe. On the other hand, the matter-dominated phase of the Universe and some final stages of stellar evolution, such as white dwarfs stars, are conveniently described in the context of Newtonian theory.

In some situations, the application of GR faces important limitations due to technical aspects. One important example are numerical simulations of structure formation in the Universe \cite{klypin, bykov}. These are a crucial approach in order to test cosmological models, mainly due to the increasing quantity of observational data on the shape and distribution of clustered matter in the form of galaxies, clusters of galaxies and filaments. In order to reproduce the observed features of these structures it is necessary to perform a deep analysis into the non-linear regime of evolution of fluctuations, taking into account many astrophysical effects such as star formation and supernovae feedback. Numerical simulations are the best approach to such problem. However, such simulations can only be performed using Newtonian physics.

This is a serious restriction because Newtonian cosmology reproduces relativistic cosmology only when the pressure is zero. Many attempts have been made in order to introduce in Newtonian cosmology pressure effects typical of the relativistic theory. One example are the so-called Neo-Newtonian hydrodynamic equations. The usual equations of Newtonian hydrodynamics in presence of the gravitational field \cite{milne1, milne2} are the following:
\begin{eqnarray}
\dot\rho + \nabla\cdot(\rho\vec v) &=& 0\;,\\
\dot{\vec v} + \vec v\cdot\nabla\vec v &=& - \frac{\nabla p}{\rho} - \nabla\phi\;,\\
\nabla^2\phi &=& 4\pi G\rho\;,
\end{eqnarray}
where the dot indicates partial derivative with respect to (wrt) time, $\rho$ is the fluid density, $\vec v$ is its flux velocity, and $\phi$ is the gravitational potential. One proposal of inclusion of pressure effects are the so-called Neo-Newtonian hydrodynamic equations, given by \cite{McCrea, Harrison, Lima:1996at, fabris, Fabris:2013psa, Fabris:2014ena, Oliveira:2014mka}:
\begin{eqnarray}
\dot\rho + \nabla\cdot(\rho\vec v) + \frac{p}{c^2}\nabla\cdot\vec v &=& 0\;,\\
\dot{\vec v} + \vec v\cdot\nabla\vec v &=& - \frac{\nabla p}{\rho + p/c^2} - \nabla\phi\;,\\
\nabla^2\phi &=& 4\pi G(\rho + 3p/c^2)\;.
\end{eqnarray}
Such proposal is based on a combination of thermodynamics considerations and pressure correction terms coming from the weak field limit of GR, made in an {\it ad hoc} way, and by interpreting $\rho + p/c^2$ as the inertial mass and $\rho + 3p/c^2$ as the gravitational mass \cite{McCrea, Harrison}. Neo-Newtonian equations reproduce the background evolution of relativistic homogeneous and isotropic cosmological models. However, at perturbative level, still in the cosmological context, these Neo-Newtonian equations only reproduce the corresponding GR equations in the large-scale regime, whereas numerical simulations are usually performed on small-scales.

Recently, using a perturbative approach in GR, Hwang and Noh \cite{Hwang:2013sia}, obtained the following set of Newtonian equations with relativistic corrections due to pressure: 
\begin{eqnarray}
\label{eh1}
\dot{\rho} + \nabla\cdot \left[(\rho + p/c^2)\vec v\right] &=& \frac{2}{c^2}\vec v\cdot\nabla p\;,\\
\label{eh2}
\dot{\vec v} + \vec v\cdot\nabla\vec v  &=&  - \nabla\phi-\frac{1}{\rho+ p/c^2}(\nabla p + \vec v\dot{p}/c^2)\;,\\
\label{eh3}
\nabla^2\phi = 4\pi G\rho\;.
\end{eqnarray}
In order to obtain these equations, the authors followed a perturbative approach about flat space-time (Minkowski space) in GR.

The goal of the present paper is to revisit the deduction of the equations proposed by Hwang and Noh. We show that a careful expansion of Einstein's equations around a flat background, keeping contributions of pressure and square velocity at first order, leads to Hwang-Noh equations. 
Moreover, we show that, while these equations seem not suitable to model the background cosmological scenario and the profile of compact objects like neutron stars (even if in the latter case some important qualitative features are reproduced), they can lead to very convenient results at perturbative level at small scales. This property may render Hwang-Noh equations interesting for applications to the analysis of structure formation in the deep sub-Hubble-scales regime.

This paper is organised as follows. In Section~\ref{sec:2}, we obtain eqs. (\ref{eh1}), (\ref{eh2}), (\ref{eh3}) from GR, using a small field and small velocity approximation scheme, and keeping the pressure correction terms. In Section~\ref{sec:3}, we exploit these equations in the cosmological context. We show that the acceleration expansion is obtained when the pressure is positive. A type Ia supernovae analysis is made for the resulting scenarios. In Section~\ref{sec:4}, a perturbative study is made around the cosmological solution, and the equations obtained in Ref.~\cite{Ma:1995ey} from a complete relativistic framework, but in the small-scales regime, are recovered. In Section~\ref{sec:5}, the corresponding equations for stellar equilibrium are obtained, and applied for the investigation of neutron stars stability. The upper mass limit for these objects turns out to be far too large than that indicated by observation, but some qualitative features of the GR case are obtained. In Section~\ref{sec:concl}, we present our final comments and conclusions.
 

\section{Pressure corrections to Newtonian hydrodynamics}\label{sec:2}

In Ref.~\cite{Hwang:2013sia}, Hwang and Noh present a set of non-relativistic hydrodynamics equations in which pressure corrections are present. In this section, we offer a derivation of such equations, inspired by Refs.~\cite{Bartelmann, Creminelli:2009mu, Carroll:2004st, Mukhanov, Schutz, Rezzolla, Poisson, Weinberg:1972kfs}. In the following sections, we apply them to cosmology and stellar equilibrium and compare their predictions with those from GR and from the Neo-Newtonian theory.
 
Let's start from GR and the following metric:
\begin{equation}\label{metric}
	ds^2 = - (1 + 2\phi/c^2)c^2dt^2 + (1 - 2\psi/c^2)\delta_{ij}dx^idx^j\;,
\end{equation}
where $\phi$ and $\psi$ are two functions of $t$ and $x^i$ (the gravitational potentials). It is useful to maintain explicit the presence of the speed of light $c$, in order to perform a post-Newtonian analysis, see e.g. \cite{Weinberg:1972kfs}. Let's also consider a perfect fluid with pressure $p$, energy density $\rho c^2$ and 4-velocity $u^\mu$. Its energy-momentum tensor can be written as:
\begin{equation}\label{enmomtens}
	T^\mu{}_{\nu} = (\rho c^2 + p)u^\mu u_\nu + p\delta^\mu{}_\nu\;.
\end{equation}
Introducing the projector 
\begin{equation}
	h^{\mu\nu} = g^{\mu\nu} + u^\mu u^\nu\;,
\end{equation}
on the 3-space orthogonal to the 4-velocity vector, we can split the energy-momentum tensor conservation
\begin{equation}
	\nabla_\mu T^\mu{}_\nu = 0\;,
\end{equation}
into the continuity and Euler equations:
\begin{eqnarray}
\label{conteq1}	u^\mu\nabla_\mu\rho c^2 +(\nabla_\mu u^\mu)(\rho c^2 + p) &=& 0\;,\\
\label{Eulereq}	(\rho c^2 + p)u^\mu\nabla_\mu u^\alpha + h^{\mu\alpha}\nabla_{\mu}p &=& 0\;.
\end{eqnarray}
The nabla denotes the covariant derivative computed wrt metric \eqref{metric}.

We investigate a post-Newtonian expansion of the above eqs.~\eqref{conteq1} and \eqref{Eulereq} for metric~\eqref{metric}. 
In particular, we shall consider: $i)$ weak fields, i.e. the gravitational potentials $\phi/c^2$ and $\psi/c^2$ are to be considered very small; $ii)$ small velocities, i.e. $dx^i/(cdt) \ll 1$. This translates also in the fact that $\partial f/\partial(ct) \ll \partial f/\partial x^i$, for a generic function $f$ of space-time, i.e. the spatial variation of a function is much larger than its time variation. These conditions can be realised by taking the $c\to\infty$ limit.

From the 4-velocity normalisation, one obtains that:
\begin{equation}\label{4velnorm}
	1 = (1 + 2\phi/c^2)(u^0)^2 - (1 - 2\psi/c^2)\delta_{ij}u^i u^j\;,
\end{equation}
where we used the definition $u^\mu \equiv dx^\mu/ds$. Define:
\begin{equation}
	u^2 \equiv (1 - 2\psi/c^2)\delta_{ij}u^i u^j = g_{ij}u^i u^j\;,
\end{equation}
i.e. the modulus of the proper 3-velocity. Note that, since $u^i \equiv dx^i/ds$, then $u^2 = \mathcal{O}(1/c^2)$. From eq.~\eqref{4velnorm}, $u^0$ can be cast in the following form:
\begin{equation}
	u^0 = 1 - \frac{\phi}{c^2} + \frac{u^2}{2} + \mathcal{O}(1/c^4)\;.
\end{equation}
We now expand eqs.~\eqref{conteq1} and \eqref{Eulereq} in powers of $1/c$, distinguishing between the cases in which $p = \mathcal{O}(c^2)$, thus a relativistic pressure, or $p = \mathcal{O}(c^0)$.

\subsection{The continuity equation}

{Assuming that $p = \mathcal{O}(c^2)$, the continuity equation~\eqref{conteq1} can be expanded in the following way:
\begin{eqnarray}\label{conteqapprox}
	\left(1 - \frac{\phi}{c^2} + \frac{u^2}{2}\right)\dot\rho c + u^i\partial_i\rho c^2 + \nonumber\\ 
	+ (\rho c^2 + p)\left[c^{-1}\left(\frac{u^2}{2}\right)^\bullet + \partial_l u^l - 3c^{-1}\frac{\dot\psi}{c^2} - 3u^l\partial_l\frac{\psi}{c^2} + u^l\partial_l\frac{\phi}{c^2}\right] + \mathcal{O}(1/c^3) = 0\;.
\end{eqnarray}}
{On the other hand, if $p = \mathcal{O}(c^0)$, eq.~\eqref{conteq1} becomes:
\begin{eqnarray}\label{conteqapprox2}
	\left(1 - \frac{\phi}{c^2} + \frac{u^2}{2}\right)\dot\rho c + u^i\partial_i\rho c^2 + (\rho c^2 + p)\partial_l u^l \nonumber\\ 
	+ \rho c^2\left[c^{-1}\left(\frac{u^2}{2}\right)^\bullet - 3c^{-1}\frac{\dot\psi}{c^2} - 3u^l\partial_l\frac{\psi}{c^2} + u^l\partial_l\frac{\phi}{c^2}\right] + \mathcal{O}(1/c^3) = 0\;.
\end{eqnarray}}
{Remarkably, the dominant $\mathcal{O}(c)$ contribution of eq.~\eqref{conteqapprox} is:
\begin{equation}
	\dot\rho c + u^i\partial_i\rho c^2 + (\rho c^2 + p)\partial_l u^l + \mathcal{O}(1/c)= 0\;,
\end{equation}
which is the well-known Newtonian result, corrected by the pressure contribution.}

\subsection{Euler equation with $\alpha = i$}

{Euler equation~\eqref{Eulereq} with $\alpha = i$ and $p = \mathcal{O}(c^2)$, can be expanded as follows:
\begin{equation}\label{Euleralphai}
	(\rho c^2 + p)\left(c^{-1}\dot{u}^i + u^l\partial_l u^i + \partial^i\frac{\phi}{c^2}\right)  + \partial^i p + u^ic^{-1}\dot{p} + u^iu^l\partial_l p + 2\frac{\psi}{c^2}\partial^ip + \mathcal{O}(1/c^2) = 0\;.
\end{equation}
Notice the following very important point. This equation has a $\mathcal{O}(c^2)$ and a $\mathcal{O}(c^0)$ contributions. The former is simply $\partial^ip = 0$, due to our assumption $p = \mathcal{O}(c^2)$. But this implies that the pressure must depend only on time! This provides a serious limitation to the use of the conservation equations with relativistic pressure contributions.}

{On the other hand, if $p = \mathcal{O}(c^0)$, Euler equation can be expanded as follows:
\begin{equation}\label{Euleralphai-bis}
	\rho c^2\left(c^{-1}\dot{u}^i + u^l\partial_l u^i + \partial^i\frac{\phi}{c^2}\right)  + \partial^i p + \mathcal{O}(1/c^2) = 0\;,
\end{equation}
reproducing the expected and well-known Newtonian result.}

\subsection{Euler equation with $\alpha = 0$}
 
{The Euler equation~\eqref{Eulereq} with $\alpha = 0$ and $p = \mathcal{O}(c^2)$ is:
\begin{equation}\label{Euleralpha0}
	\left(1 - \frac{\phi}{c^2} + \frac{u^2}{2}\right)u^l\partial_lp + u^2\frac{1}{c}\dot{p} + (\rho c^2 + p)\left[\frac{1}{c}\left(\frac{u^2}{2}\right)^\bullet + u^l\partial_l\frac{\phi}{c^2} + u^l\partial_l\left(\frac{u^2}{2}\right)\right] + \mathcal{O}(1/c^3) = 0\;.
\end{equation}
Again, we retained in the above equation $\mathcal{O}(c)$ and $\mathcal{O}(1/c)$ contributions. The former is very simple, i.e. $u^l\partial_lp = 0$, and gives us a constraint which is compatible with $\partial_lp = 0$ we found earlier.}  

{The Euler equation~\eqref{Eulereq} with $\alpha = 0$ and $p = \mathcal{O}(c^0)$ is:
\begin{equation}\label{Euleralpha0bis}
	u^l\partial_lp + \rho c\left(\frac{u^2}{2}\right)^\bullet + u^l\rho\partial_l\phi + \rho c^2u^l\partial_l\left(\frac{u^2}{2}\right) + \mathcal{O}(1/c^3) = 0\;.
\end{equation}
Note here the absence of the $\dot{p}$ term wrt eq.~\eqref{Euleralpha0}. This fact will prove to be crucial when we will investigate cosmology using the Hwang-Noh equations.}  

\subsection{The Hwang-Noh equations}

{We now try and understand how to derive the Hwang-Noh equations from the above expansions of the continuity and Euler equations. The Hwang-Noh equations are the following \cite{Hwang:2013sia}:
\begin{eqnarray}
\label{newconteq}\dot{\rho}c + \partial_l\left[(\rho c^2 + p)u^l\right] = 2u^l\partial_lp\;,\\
\label{newEuler} c^{-1}\dot{u}^i + u^l\partial_l u^i + \partial^i\frac{\phi}{c^2} = -\frac{1}{\rho c^2 + p}(\partial^i p + u^ic^{-1}\dot{p})\;.
\end{eqnarray}
It seems clear that, in order to reproduce the Euler equation \eqref{newEuler}, we must indeed assume that $p = \mathcal{O}(c^2)$ in order to have the $\rho c^2 + p$ term at the denominator of the right hand side (rhs). This is already problematic in itself, because we saw that in doing so the pressure must be homogeneous and isotropic. Moreover, in order to obtain eq.~\eqref{newEuler} from eq.~\eqref{Euleralphai} we must drop for some reason the contributions $u^iu^l\partial_l p + 2\psi\partial_ip/c^2$, which are also of order $\mathcal{O}(c^0)$. Neglecting these terms could be justified in cosmological perturbations theory, if we assume $p \to \delta p$ and $\psi \to \delta\psi$ and thus neglect second order terms. It is presumably for this reason that the Hwang-Noh equations work well in this regime, as we show in Sec.~\ref{sec:4}.}

{Justifying eq.~\eqref{newconteq} is easier. Rewrite eq.~\eqref{conteqapprox} as follows, by multiplying it by $1 + \phi/c^2 - u^2/2$:
\begin{eqnarray}\label{conteqapprox-approx}
	\dot\rho c + \partial_l\left[(\rho c^2 + p)u^l\right] = u^l\partial_lp\left(1 + \frac{\phi}{c^2} - \frac{u^2}{2}\right) \nonumber\\ - (\rho c^2 + p)\left[c^{-1}\left(\frac{u^2}{2}\right)^\bullet - 3c^{-1}\frac{\dot\psi}{c^2} - 3u^l\partial_l\frac{\psi}{c^2} + u^l\partial_l\frac{\phi}{c^2}\right] + \mathcal{O}(1/c^3)\;,
\end{eqnarray}
and again neglecting $\mathcal{O}(1/c^3)$ terms. If we now sum this equation to eq.~\eqref{Euleralpha0} and neglect $\mathcal{O}(1/c)$ term, we find eq.~\eqref{newconteq}. Of course, there is a trick hidden here. We showed right after eq.~\eqref{Euleralpha0} that if $p = \mathcal{O}(c^2)$ then $u^l\partial_lp = 0$. Therefore, whatever factor multiplying $u^l\partial_lp$ in eq.~\eqref{newconteq} is good, including 2.}

%

We derive now the Poisson equation found in \cite{Hwang:2013sia}, i.e. $\nabla^2\phi = 4\pi G\rho$.

\subsection{Gravitational field equations}

{The Einstein tensor components are easily calculated from metric~\eqref{metric} and are the following, to the lowest order in the $1/c$ expansion (see also \cite{Weinberg:1972kfs}):
\begin{eqnarray}
	G_{00} &=& 2\nabla^2\frac{\psi}{c^2} + \mathcal{O}(1/c^4)\;,\\ \
	G_{0i} &=& 2c^{-1}\frac{\dot\psi_{,i}}{c^2} + \mathcal{O}(1/c^5)\;, \\ 
\label{Gijeq}	G_{ij} &=& \frac{1}{c^2}\nabla^2(\phi - \psi)\delta_{ij} + \frac{1}{c^2}(\psi - \phi)_{,ij} + \mathcal{O}(1/c^4)\;.
\end{eqnarray}}
We must couple this Einstein tensor to the energy-momentum tensor~\eqref{enmomtens}, also expanded at the appropriate order in powers of $1/c$. We get the following Einstein equations. The $0-0$ one is:
\begin{equation}\label{Poisseq}
	\nabla^2\psi = 4\pi G\rho\;. \qquad \mathcal{O}(1/c^2)\;.
\end{equation}
This Poisson equation is also consistent with Einstein's equations projected along the 4-velocity, i.e. $G_{\mu\nu}u^\mu u^\nu = (8\pi G/c^4)T_{\mu\nu}u^\mu u^\nu$.
 
The $0-i$ Einstein's equation is the following:
\begin{equation}
	c^{-1}\frac{\dot\psi_{,i}}{c^2} = -\frac{4\pi G}{c^4}(\rho c^2 + p)u_i\;, \qquad \mathcal{O}(1/c^3)\;.
\end{equation}
{In the static case, i.e. $\dot\psi = 0$, one would obtain $u_i = 0$. Therefore, staticity seems to be inconsistent with a fluid flow. On the other hand, the above equation is of $\mathcal{O}(1/c^3)$ order, thus if we truncate the overall expansion of our theory at $\mathcal{O}(1/c^2)$, it is still consistent to have static potentials, as assumed e.g. in \cite{Bartelmann}.} 

{The spatial trace of the Einstein equation $i-j$ gives:
\begin{equation}\label{spatialtraceEinsteinEq}
	 \frac{1}{c^2}\nabla^2(\phi - \psi) = \frac{12\pi G}{c^4}p\;, \qquad \mathcal{O}(1/c^2)\;,
\end{equation}
whereas the traceless spatial Einstein's equation can be cast in the following way:
\begin{equation}\label{spatialtracelessEinsteq}
	\frac{1}{c^2}\left(\frac{1}{3}\delta_{ij}\nabla^2 - \nabla_i\nabla_j\right)(\phi - \psi) = 0\;, \qquad \mathcal{O}(1/c^2)\;.
\end{equation}
We must look carefully at eq.~\eqref{spatialtraceEinsteinEq}: it holds true only if $p = \mathcal{O}(c^2)$. Indeed, in this case $p/c^4 = \mathcal{O}(1/c^2)$ and therefore the left hand side (lhs) and the rhs of eq.~\eqref{spatialtraceEinsteinEq} are consistent. If this is the case, we then combine eq.~\eqref{spatialtraceEinsteinEq} with Poisson equation \eqref{Poisseq} and obtain the well-known result $\nabla^2\phi = 4\pi G(\rho + 3p/c^2)$, which is taken into account e.g. in the neo-Newtonian theory. This is a result of interest e.g. in cosmology, where the equation of state $p = w\rho c^2$ is often used.}

{However, the Poisson equation found in \cite{Hwang:2013sia} has no pressure contribution: 
\begin{equation}\label{HwangPoisseq}
	\nabla^2\phi = 4\pi G\rho\;.
\end{equation}
We can reproduce the latter if we drop the assumption $p = \mathcal{O}(c^2)$. In this case the rhs of eq.~\eqref{spatialtraceEinsteinEq} is of order $\mathcal{O}(1/c^4)$ and is inconsistent with the lhs. Therefore, we can assume $\phi = \psi$, a condition which is also compatible with eq.~\eqref{spatialtracelessEinsteq}. However, if we drop the assumption $p = \mathcal{O}(c^2)$, then eqs.~\eqref{newconteq} and \eqref{newEuler} are no more justified.} 

{Forcing the use the ansatz $p = w\rho c^2$ in the Hwang-Noh equations reflects in a weird cosmological behaviour, as we show in the next section.}

\section{Newtonian cosmology with pressure corrections}\label{sec:3}

In this section we apply eqs.~\eqref{newconteq} and \eqref{newEuler} to cosmology. We assume that density and pressure depend only on time and that $u^i = H(t)x^i/c$, in order to simulate the Hubble flux. 
Equations \eqref{newconteq} and \eqref{newEuler} become:
\begin{eqnarray}
\label{conteq}	\dot{\rho}c^2 + 3H(\rho c^2 + p) &=& 0\;,\\
\label{Euleqcosmo}	c^{-2}\dot{H}x^i + c^{-2}H^2x^i + \partial^i\frac{\phi}{c^2} &=& -\frac{1}{\rho c^2 + p}Hx^ic^{-2}\dot{p}\;.
\end{eqnarray}
The first equation~\eqref{conteq} is identical to the continuity equation of relativistic cosmology, see e.g. Ref.~\cite{Mukhanov}. Taking the divergence of the second equation and using Poisson equation~\eqref{HwangPoisseq}, we obtain:
\begin{equation}\label{corrHdoteq}
	\dot{H} + H^2 = -\frac{4\pi G}{3}\rho -\frac{1}{\rho c^2 + p}H\dot{p}\;. \qquad \mbox{(Hwang-Noh)}
	\end{equation}
This equation is different from the relativistic one, which is the following:
\begin{equation}\label{GRHdoteq}
	\dot{H} + H^2 = -\frac{4\pi G}{3}\rho - 4\pi G \frac{p}{c^2}\;, \qquad \mbox{(GR)}
\end{equation}
except, of course, for the dust ($p = 0$) case and for a specific case in which the pressure satisfies the equation
\begin{equation}\label{presseq}
	H\dot p = 4\pi G\frac{p}{c^2}(\rho c^2 + p)\;,
\end{equation}
obtained by equating eqs.~\eqref{corrHdoteq} and \eqref{GRHdoteq}. For this last possibility, in the barotropic case $p = w\rho c^2$, with $w =$ constant, one obtains from eq.~\eqref{presseq} and eq.~\eqref{conteq} that
\begin{equation}
	-3H^2 = 4\pi G\rho\;,
\end{equation}
which is inconsistent. Considering instead a time-dependent $w$, one obtains the following equation for $H$:
\begin{equation}
	H\frac{\dot{w}}{w(1 + w)} - 3H^2 = 4\pi G\rho\;.
\end{equation}
An interesting case is that for a politropic equation of state, such as $p = \mathcal{K}(\rho c^2)^\gamma$, where $\mathcal{K}$ is some constant with suitable units of measure and $\gamma$ is the politropic index. Substituting this ansatz into eq.~\eqref{presseq}, and using the continuity equation~\eqref{conteq}, one obtains:
\begin{equation}
	H^2 = -\frac{1}{2\gamma}\frac{8\pi G}{3}\rho\;.
\end{equation}
This Friedmann equation makes sense only if $\gamma < 0$, which turns the fluid in the famous generalised Chaplygin gas \cite{Kamenshchik:2001cp}, which naturally describes the transition from a decelerated phase of expansion of the Universe to an accelerated one.

We come back and focus our attention on the general case of eq.~\eqref{corrHdoteq}, assuming again a barotropic equation of state $p = w\rho c^2$ and using eq.~\eqref{conteq}:
\begin{equation}\label{corrHdoteqbaro}
	\dot{H} + (1 - 3w)H^2 = -\frac{4\pi G}{3}\rho\;.
\end{equation}
The acceleration of the expansion is given by:
\begin{equation}\label{acceq}
	\frac{\ddot{a}}{a} = 3wH^2 - \frac{4\pi G}{3}\rho\;.
\end{equation}
Therefore, a necessary condition in order to have an accelerated expansion is that $w > 0$, in contrast with GR, which demands that $w < -1/3$. In particular, we can predict an accelerated expansion of the Universe without invoking Dark Energy! {We come back on this fundamental issue at the end of this section. Now we investigate in some detail the conditions in order to have an accelerated expansion of the Universe using Hwang-Noh equations.}

Considering the case of constant $w$, the continuity equation~\eqref{conteq} gives the known result
\begin{equation}
	\rho = \rho_0a^{-3(1 + w)}\;,
\end{equation}
and eq.~\eqref{corrHdoteqbaro} can be solved exactly, using the scale factor as independent variable:
\begin{equation}\label{Friedeq1}
	H^2 = \frac{8\pi G}{3(1 + 9w)}\rho_0a^{-3(1 + w)} + \frac{K}{a^{2(1 - 3w)}}\;, \qquad w \neq -1/9\;,
\end{equation}
where $K$ is an integration constant, recalling the spatial curvature contribution of relativistic cosmology. Compared with GR, the gravitational constant is effectively changed $G \to G/(1 + 9w)$ and the curvature term has a different evolution. Depending on the sign of $K$ and on the values of $w$, we have many possible evolutions.

Combining eqs.~\eqref{acceq} and \eqref{Friedeq1}, we find the following condition for the accelerated expansion:
\begin{equation}\label{acccond}
	wKa^{(1 + 9w)} > \frac{4\pi G\rho_0(1 + 3w)}{9(1 + 9w)}\;.
\end{equation}
If $w > 0$ and $K > 0$, there exists an accelerated phase of expansion, starting from the threshold scale factor:
\begin{equation}\label{atrans}
	a_{\rm T} = \left[\frac{4\pi G\rho_0(1 + 3w)}{9wK(1 + 9w)}\right]^{\frac{1}{1 + 9w}}\;.
\end{equation} 
For $w = 0$ or $K = 0$, the above threshold scale factor diverges, leaving thus the Universe always in a decelerated phase of expansion. It is interesting to notice that, if we associate $K$ to the spatial curvature as we do in GR, then it is its presence that guarantees an accelerated phase of expansion, differently from GR.

If $w = -1/9$, the solution for $H^2$ is logarithmic:
\begin{equation}
	H^2a^{8/3} = K - \frac{8\pi G}{3}\rho_0\ln a\;.
\end{equation} 
This evolution is always limited to the interval $0 < a < a_F$, where $a_F$ is the final scale factor, for which $H^2$ changes sign becoming negative. Indeed, in order to have $H^2 > 0$, one has to demand that:
\begin{equation}
	\ln a < \frac{3K}{8\pi G\rho_0} \equiv \ln a_F\;.
\end{equation}
Since $w = -1/9$, the expansion is always decelerated.

\begin{figure}[htbp]
\centering
	\includegraphics[width=0.7\columnwidth]{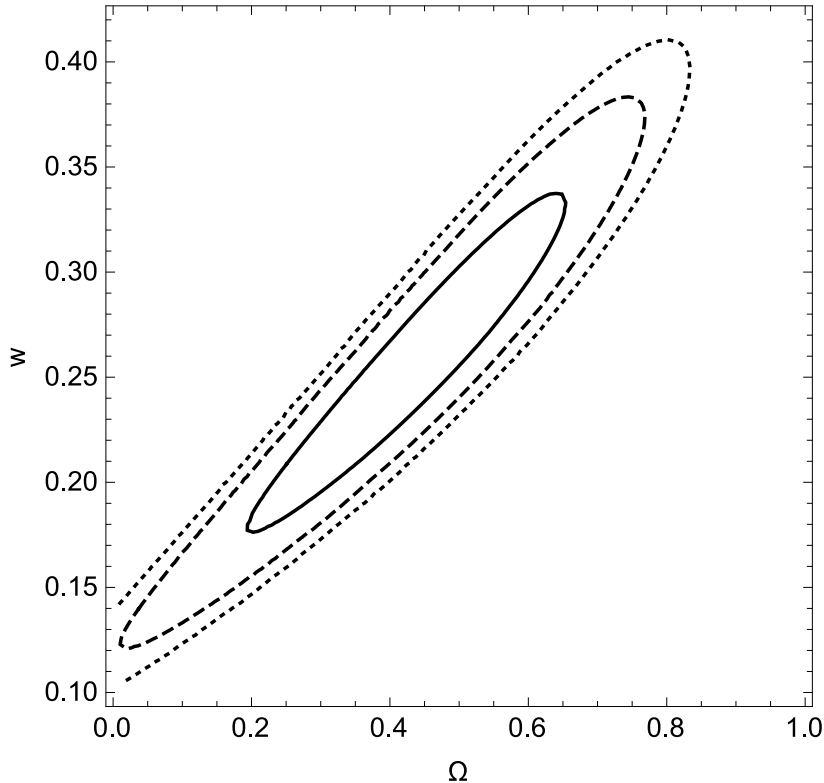}
	\caption{Contour plots in the parameter space $\Omega-w$ at 66\%, 95\% and 99\% confidence level.}
	\label{Fig:contplot}
\end{figure}

In order to see what data have to say about Friedmann equation~\eqref{Friedeq1} we perform a Bayesian analysis using Union2 type Ia Supernovae dataset \cite{Suzuki:2011hu}. We define, as usual,
\begin{equation}
	\Omega \equiv \frac{8\pi G\rho_0}{3H_0^2}\;,
\end{equation}
where $H_0$ is the Hubble constant. {Friedmann's eq.~\eqref{Friedeq1} thus becomes:
\begin{equation}\label{Friedeq1bis}
	\frac{H^2}{H_0^2} = \frac{\Omega}{(1 + 9w)}a^{-3(1 + w)} + \frac{1 - \frac{\Omega}{1 + 9w}}{a^{2(1 - 3w)}}\;, \qquad w \neq -1/9\;.
\end{equation}}
In Fig.~\ref{Fig:contplot} we show the posterior probability contour plots for 66\%, 95\% and 99\% confidence level in the $\Omega-w$ plane. {The analysis gives a $\chi^2 = 562.336$ ($\chi_{\rm red} = 0.969545$) for $\Omega = 0.46^{+0.27}_{-0.33}$ and $w = 0.26^{+0.10}_{-0.11}$ at 95\% level, upon marginalisation. From eq.~\eqref{atrans} we can compute the transition scale factor as $a_{\rm T} = 0.60$, which corresponds to a redshift $z_{\rm T} = 0.66$.} 

{The same analysis performed in GR for a $w$CDM model, i.e. for the following Friedmann equation:
\begin{equation}\label{FriedeqwCDM}
	\frac{H^2}{H_0^2} = \Omega a^{-3} + (1 - \Omega)a^{-3(1 + w)}\;,
\end{equation}
gives a $\chi^2 = 562.225$ ($\chi_{\rm red} = 0.969935$) for $\Omega = 0.30^{+0.10}_{-0.18}$ and $w = -0.99^{+0.33}_{-0.40}$ at 95\% level, upon marginalisation. As expected, data point towards the $\Lambda$CDM model. With these results, we can compute the scale factor at the transition to the accelerated phase: $a_{\rm T} = 0.60$, which corresponds to the redshift $z_{\rm T} = 0.67$. Remarkably, Hwang-Noh equations and GR are equally successful in fitting type Ia Supernovae data despite their contradiction on the value of $w$ required in order to generate an accelerated expansion.}

{How could we explain this contradiction? After all, we derived Hwang-Noh equations starting from GR. We try and identify the issue in the following.}

{The issue of generating an accelerated expansion with a positive pressure is due to the last term of \eqref{corrHdoteq},	i.e. the term proportional to $\dot p$. In turn, this term can be traced back to the generalised Euler equation \eqref{Euleralphai} and it is absent in the usual Newtonian theory, i.e. in eq.~\eqref{Euleralphai-bis}. However, this term is also present in the time-component of the Euler equation, i.e. eq.~\eqref{Euleralpha0}. Therefore, this equation also provides an evolution equation for $p$ and recall that this happens since we assumed $p = \mathcal{O}(c^2)$. Equation~\eqref{Euleralpha0}, together with $\partial_lp = 0$ gives the following equation:
\begin{equation}
	\dot{p} + (\rho c^2 +p)\left(\frac{\dot{H}}{H} + H\right) = 0\;.
\end{equation}
Combining this equation with eq.~\eqref{conteq} and the equation of state ansatz $p = w\rho c^2$ results in the following:
\begin{equation}
	\dot{H} + (1 - 3w)H^2 = 0\;.
\end{equation}
This equation, together with eq.~\eqref{corrHdoteq}, implies that $\rho = 0$, i.e. absence of a matter content. Therefore, the contradiction with GR that we found earlier is due to an incompatibility among the cosmological ansatz, the equation of state $p = w\rho c^2$ and Euler and the continuity equations.} 

%

{We stress that the above mentioned inconsistence is not an issue of Hwang-Noh theory only. Consider for example the hydrodynamics equations derived in \cite{Bartelmann}, i.e. eqs.~(3.106) of this reference. Together with the cosmological ansatz $u^i =Hx^i/c$, we get the following equation instead of eq.~\eqref{corrHdoteq}:
\begin{equation}
	\dot{H} + H^2 = -\frac{4\pi G}{3}\rho(1 + 3w) -\frac{H\dot{p}}{\rho c^2 + p}\;.
	\end{equation}
The only difference from the Hwang-Noh case is the $(1 + 3w)$ factor multiplying $\rho$, coming from the the $3p$ contribution of the Poisson equation used in \cite{Bartelmann}, which is absent in eq.~\eqref{HwangPoisseq}.} 
{From the above equation we get the solution 
\begin{equation}\label{BartFriedeq}
	H^2 = \frac{8\pi G(1 + 3w)}{3(1 + 9w)}\rho_0a^{-3(1 + w)} + \frac{K}{a^{2(1 - 3w)}}\;, \qquad w \neq -1/9\;,
\end{equation}
which should be compared with equation \eqref{Friedeq1}, derived from Hwang-Noh equations. From eq.~\eqref{BartFriedeq} we also find for $w > 0$ and $K > 0$ a transition to an accelerated phase of expansion, given by:
\begin{equation}
	wKa^{(1 + 9w)} > \frac{4\pi G\rho_0(1 + 3w)^2}{9(1 + 9w)}\;.
\end{equation}
It has an extra $1 + 3w$ factor which does not avoid the transition to an accelerated phase of expansion, but makes it happen later wrt the condition given in eq.~\eqref{acccond}.}

\section{Evolution of small cosmological perturbations}\label{sec:4}

In this section we consider fluctuations about the cosmological solution given by eqs.~\eqref{conteq} and \eqref{Euleqcosmo}. To this purpose, we introduce fluctuations in the physical quantities in the following way:
\begin{eqnarray}
	\epsilon = \epsilon_0 + \delta\epsilon\;, \quad p = p_0 + \delta p &=& p_0 + c_s^2\delta\epsilon\;, \quad u^i = Hx^i/c + \delta u^i\;, \quad \phi = \phi_0 + \delta\phi\;,
\end{eqnarray}
where we defined $\epsilon_0 = \rho_0c^2$. We assumed here that $\delta p = c_s^2\delta\epsilon$, i.e. the perturbation is adiabatic; $c_s^2$ is the adiabatic speed of sound and we assume it to be constant. {Moreover, we assume $p_0 = w\epsilon_0$, with $w$ also constant. This implies $w = c_s^2$, but we shall maintain the two quantities separated in order to better show how the structure of the equations found resembles that of GR.}

With these positions, fluctuations of eqs.~\eqref{newconteq} and \eqref{newEuler} give:
\begin{eqnarray}
	\dot{\delta\epsilon} + \partial_l\left[(\epsilon_0 + p_0)c\delta u^l + \delta\epsilon(1 + c_s^2)Hx^l\right] = 2Hx^l\partial_l\delta p\;,\\
	\dot{\delta u^i} + (Hx^l\partial_l)\delta u^i + (\delta u^l\partial_l)(Hx^i) + c^{-1}\partial^i\delta\phi =\nonumber\\ -\frac{1}{\epsilon_0 + p _0}(c\partial^i\delta p + \dot{p_0}\delta u^i + Hx^i\dot{\delta p}/c) + \frac{\delta\epsilon + \delta p}{(\epsilon_0 + p_0)^2}Hx^i\dot{p_0}/c\;.
\end{eqnarray}
Introducing the density contrast $\delta \equiv \delta\epsilon/\epsilon_0$, one can cast the above equations as follows:
\begin{eqnarray}\label{deltaeq1}
	\dot{\delta} + c(1 + w)\partial_l\delta u^l + 3H\delta(c_s^2 - w) + Hx^l(1 - c_s^2)\partial_l\delta = 0\;,\\
\label{deltau1}	\dot{\delta u^i} + (Hx^l\partial_l)\delta u^i + H\delta u^i + c^{-1}\partial^i\delta\phi =\nonumber\\ -\frac{1}{1 + w}\left(cc_s^2\partial^i\delta + \frac{\dot{p_0}}{\epsilon_0}\delta u^i + Hx^i\frac{\dot{\delta p}}{c\epsilon_0}\right) + \delta\frac{1 + c_s^2}{(1 + w)^2}Hx^i\frac{\dot{p_0}}{c\epsilon_0}\;.
\end{eqnarray}
Unfortunately, here appears the problem of having $x^i$ in this equations. This issue is solved in the standard Newtonian calculations by transforming to Lagrangian coordinates, i.e. $x^i = aq^i$. See e.g. Ref.~\cite{Mukhanov}. {In particular, one has
\begin{equation}
	\left.\frac{\partial}{\partial t}\right|_{\textbf{x}} + u^i\frac{\partial}{\partial x^i} = \left.\frac{\partial}{\partial t}\right|_{\textbf{q}}\;,
\end{equation}
which essentially transforms the term containing $x^i$ in a partial time derivative.} Performing this transformation for eqs.~\eqref{deltaeq1} and \eqref{deltau1}, we end with:
\begin{eqnarray}
	\dot{\delta} &+& c(1 + w)\frac{1}{a}\partial_l\delta u^l + 3H\delta(c_s^2 - w) - Hq^lc_s^2\partial_l\delta = 0\;,\\
	\dot{\delta u^i} &+& H\delta u^i + \frac{\partial^i\delta\phi}{ac} +\frac{1}{1 + w}\left(c\frac{c_s^2}{a}\partial^i\delta + \frac{\dot{p_0}}{\epsilon_0}\delta u^i + Haq^i\frac{\dot{\delta p}}{c\epsilon_0}\right) = \frac{1 + c_s^2}{(1 + w)^2}\frac{Haq^i\dot{p_0}\delta}{c\epsilon_0}\;.
\end{eqnarray}
Changing the time derivative of the term $Hx^i\dot\delta p$ would also generate a term proportional to $H^2a^2q^iq^l\partial_l\delta p$, which we neglected as being at least of third order. 

Therefore, the problem of having the position coordinate, now $q^i$, free in the equations still remains. {Notice that such problem also appears in the first formulation of Neo-Newtonian cosmology, see Refs.~\cite{McCrea, Lima:1996at}}. How could we solve it?

One possibility is to simply remove these terms, focusing our attention only on small scales, for which $Hq^i/c \ll 1$ and retaining the dominant terms only. If we do that, and call $\theta \equiv c\partial_l\delta u^l$, the above equations become:
\begin{eqnarray}
	\dot\delta + 3H\delta(c_s^2 - w) + (1 + w)\theta/a = 0\;,\\
	\dot\theta + H\theta(1 - 3w) = -\frac{1}{a}\nabla^2\delta\phi - \frac{c_s^2}{1 + w}\frac{1}{a}c^2\nabla^2\delta\;.
\end{eqnarray}
These are the same evolution equations for small fluctuations in relativistic cosmology, in the Newtonian gauge, see e.g. Ref.~\cite{Ma:1995ey}.\footnote{Actually the GR equations have the term $\dot\psi$, see Ref. \cite{Ma:1995ey}, which is however negligible in our approximation.} This may be not surprising, since we started from GR with the same gauge, but we had to make the approximation of small scales.

\section{Newtonian stellar equilibrium with pressure corrections and comparison with Neo-Newtonian theory and General Relativity}\label{sec:5}

In this section we employ eqs.~\eqref{newconteq} and \eqref{newEuler} in order to study the equilibrium of a pure neutron star with nuclear interactions. We compare the results with: $i)$ standard Newtonian theory, $ii)$ Neo-Newtonian theory and $iii)$ GR. Assuming spherical symmetry and staticity for eq.~\eqref{newEuler} we get:
\begin{equation}
	\frac{dp(r)}{dr} = -[\rho(r) + p(r)/c^2]\frac{d\phi(r)}{dr}\;.
\end{equation}
Any function shall be function of $r$ only from now on, so we drop the explicit functional dependence. From Poisson equation~\eqref{HwangPoisseq} we obtain
\begin{equation}
	r^2\frac{d\phi}{dr} = 4\pi G\int^r_0dr'r'^2\rho \equiv GM\;,
\end{equation}
where in the latter equation we defined the mass $M$ of the star. Thus, we obtain the following couple of equations describing the equilibrium of a spherically symmetric distribution of matter:
\begin{equation}
	\frac{dp}{dr} = -\frac{GM\rho}{r^2}\left(1 + \frac{p}{\rho c^2}\right)\;, \qquad \frac{dM}{dr} = 4\pi\rho r^2\;. \qquad \mbox{(Hwang-Noh)}
\end{equation}
These equations provide the correction $p/\rho c^2$ with respect to the Newtonian counterpart, which is the following:
\begin{equation}
	\frac{dp}{dr} = -\frac{GM\rho}{r^2}\;, \qquad \frac{dM}{dr} = 4\pi\rho r^2\;. \qquad \mbox{(Newtonian)}
\end{equation}
In the Neo-Newtonian theory Poisson's equation gets the $3p$ correction, i.e. $\nabla^2\phi = 4\pi G(\rho + 3p/c^2)$. Therefore, the stability equations are the following:
\begin{equation}
	\frac{dp}{dr} = -\frac{GM\rho}{r^2}\left(1 + \frac{p}{\rho c^2}\right)\;, \qquad \frac{dM}{dr} = 4\pi\rho r^2\left(1 + \frac{3p}{\rho c^2}\right)\;. \qquad \mbox{(Neo-Newtonian)}
\end{equation}
Note that the equation for the pressure is the same as in Hwang-Noh, but the mass equation is corrected by a term $3p/\rho c^2$. 

Finally, starting from a static, spherically symmetric metric the Tolman-Oppenheimer-Volkoff (TOV) equations are the following:
\begin{eqnarray}
	\frac{dp}{dr} = -\frac{GM\rho}{r^2}\left(1 + \frac{p}{\rho c^2}\right)\left(1 + \frac{4\pi r^3p}{Mc^2}\right)\left(1 - \frac{2GM}{c^2r}\right)^{-1}\;,\\ 
	\frac{dM}{dr} = 4\pi\rho r^2\;. \qquad \mbox{(TOV)}
\end{eqnarray}
As equation of state, we adopt the simple fit of \cite{Silbar:2003wm} for the case of a pure neutron star with nuclear interactions:
\begin{equation}
	p = \kappa_0(\rho c^2)^2\;, \quad \mbox{with} \quad \kappa_0 = 4.012\times 10^{-4}\;\mbox{fm}^3/\mbox{MeV}\;.
\end{equation}
We introduce the following dimensionless quantities:
\begin{equation}
	\tilde{r} \equiv \frac{c^2r}{GM_\odot}\;, \quad \tilde{M} \equiv \frac{M}{M_\odot}\;, \quad \tilde{\epsilon} \equiv \kappa_0\rho c^2\;, 
\end{equation}
where $M_\odot$ is our Sun's mass. With these definitions the Newtonian equations become:
\begin{equation}
	\frac{d\tilde{\epsilon}}{d\tilde{r}} = -\frac{\tilde{M}}{2\tilde{r}^2}\;, \qquad \frac{d\tilde{M}}{d\tilde{r}} = A\tilde{\epsilon}\tilde{r}^2\;, \qquad \mbox{(Newtonian)}
\end{equation}
where
\begin{equation}
	A \equiv \frac{4\pi G^3M^2_\odot}{\kappa_0c^8} = 9.066\times 10^{-2}\;.
\end{equation}
The above equations for the Newtonian case can be solved exactly. Deriving the first equation and combining it with the second, we can find the following second order equation for $\tilde{\epsilon}$:
\begin{equation}\label{Newtcase}
	\tilde{\epsilon}'' + \frac{2}{\tilde{r}}\tilde{\epsilon}' + \frac{A}{2}\tilde{\epsilon} = 0\;, \qquad \mbox{(Newtonian)}
\end{equation}
where the prime denotes derivation wrt $\tilde{r}$. Denoting as $\tilde\epsilon_0$ as the central normalized energy density, the solution of eq.~\eqref{Newtcase} is a cardinal sine function:
\begin{equation}
	\tilde{\epsilon} = \tilde\epsilon_0\frac{\sin\left(\sqrt{A/2}\tilde{r}\right)}{\sqrt{A/2}\tilde{r}}\;. \qquad \mbox{(Newtonian)}
\end{equation}
Assuming that the first zero determines the star's radius, we have:
\begin{equation}
	R_* = \sqrt{\frac{2}{A}}\frac{GM_\odot\pi}{c^2} = 22.87\;\mbox{km}\;. \qquad \mbox{(Newtonian)}
\end{equation}
This radius does not depend on the star's central density, so in Newtonian physics a pure neutron star with nuclear interactions is always stable and always has the same radius, independently from its mass.

Hwang-Noh, Neo-Newtonian and TOV equations, using the dimensionless quantities previously introduced, become:
\begin{eqnarray}
	\frac{d\tilde{\epsilon}}{d\tilde{r}} = -\frac{\tilde{M}}{2\tilde{r}^2}(1 + \tilde{\epsilon})\;, \qquad \frac{d\tilde{M}}{d\tilde{r}} = A\tilde{\epsilon}\tilde{r}^2\;, \qquad \mbox{(Hwang-Noh)}\\
	\frac{d\tilde{\epsilon}}{d\tilde{r}} = -\frac{\tilde{M}}{2\tilde{r}^2}(1 + \tilde{\epsilon})\;, \qquad \frac{d\tilde{M}}{d\tilde{r}} = A\tilde{\epsilon}\tilde{r}^2(1 + 3\tilde\epsilon)\;, \qquad \mbox{(Neo-Newtonian)}\\
	\frac{d\tilde{\epsilon}}{d\tilde{r}} = -\frac{\tilde{M}}{2\tilde{r}^2}(1 + \tilde{\epsilon})\left(1 + A\frac{\tilde{r}^3\tilde\epsilon^2}{\tilde{M}}\right)\left(1 - \frac{2\tilde{M}}{\tilde{r}}\right)^{-1}\;, \qquad \frac{d\tilde{M}}{d\tilde{r}} = A\tilde{\epsilon}\tilde{r}^2\;. \qquad \mbox{(TOV)}
\end{eqnarray}
We numerically solve these equations using as boundary conditions $\tilde\epsilon(r=0) = \tilde\epsilon_0$ and $M(r = 0) = 0$. We determine as the star's radius the one for which the energy density becomes negative and then we plot the stability curves of the masses as functions of the radii.	
\begin{figure}[htbp]
\centering
	\includegraphics[width=0.7\columnwidth]{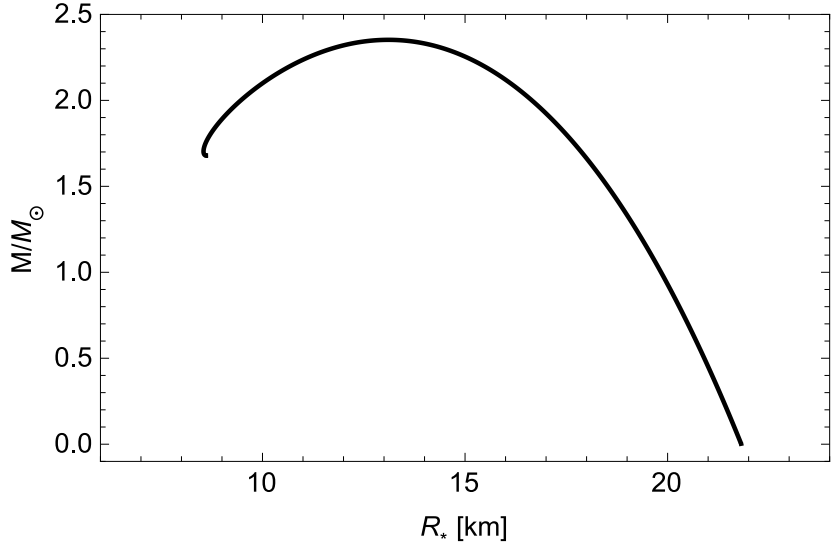}
	\caption{Stability curve for the TOV case}
	\label{Fig:StabTOV}
\end{figure}
In Fig.~\ref{Fig:StabTOV} we show the results for the TOV case, with a maximum mass below 2.5 solar masses.
\begin{figure}[htbp]
\centering
	\includegraphics[width=0.7\columnwidth]{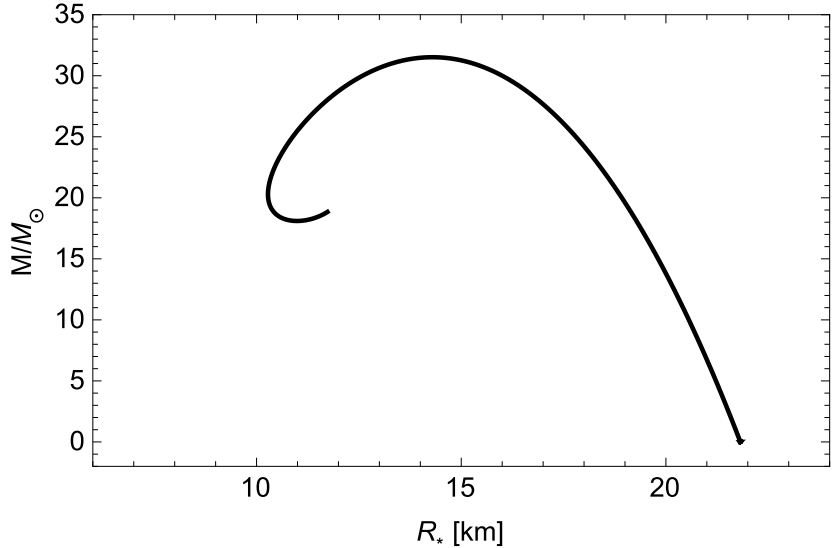}
	\caption{Stability curve for the Hwang-Noh case}
	\label{Fig:StabHN}
\end{figure}
In the Hwang-Noh case in Fig.~\ref{Fig:StabHN}, the stability curve attains a maximum mass of about 32 solar masses!
\begin{figure}[htbp]
\centering
	\includegraphics[width=0.7\columnwidth]{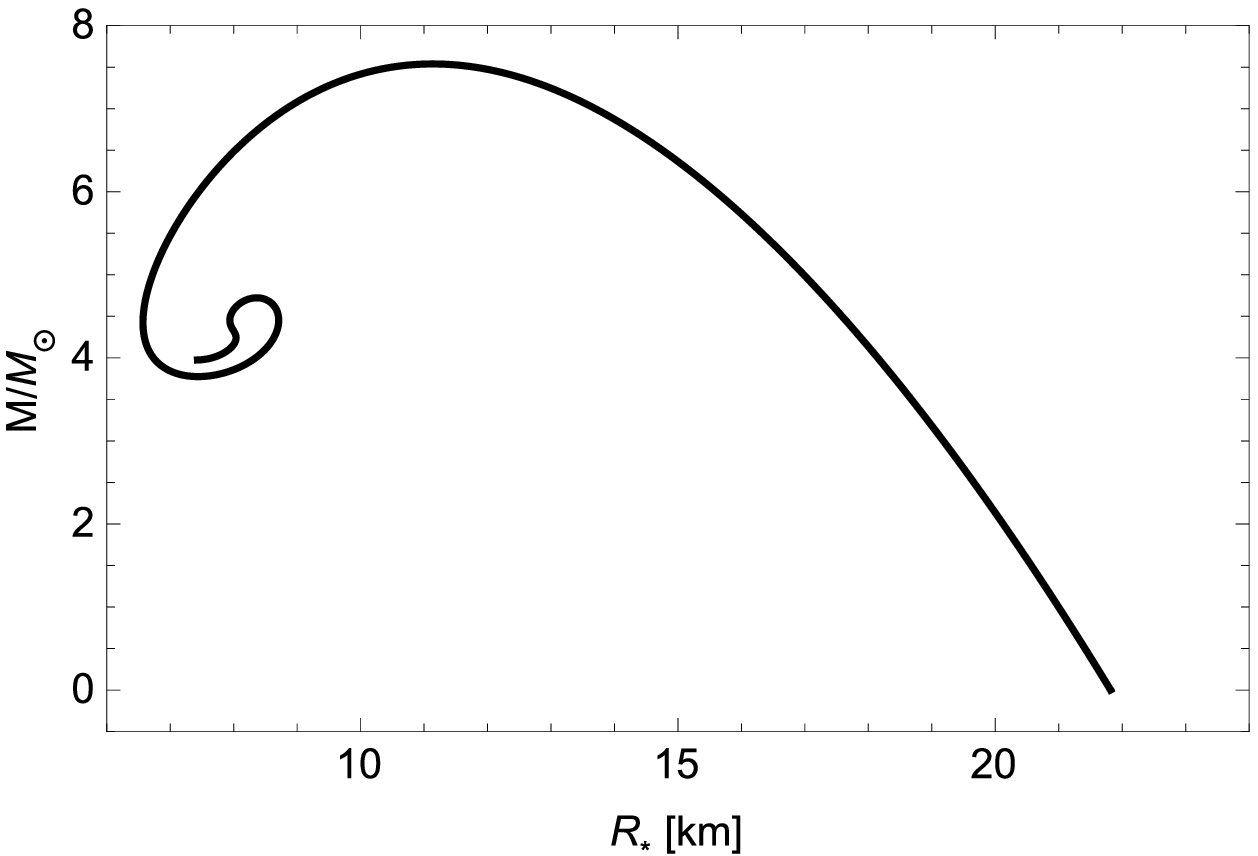}
	\caption{Stability curve for the Neo-Newtonian case}
	\label{Fig:StabNN}
\end{figure}
The Neo-Newtonian case, in Fig.~\ref{Fig:StabNN}, is the one which less deviate from the general relativistic result, predicting a maximum mass of the order of 7.5 solar masses.\footnote{In Ref. \cite{Oliveira:2014mka} other possible definitions of mass were exploited, leading to smaller values for the maximum mass allowed.}{Note that the set of equations presented in \cite{Bartelmann} are in this context identical to the ones of the Neo-Newtonian case.}

A comment is in order here. The results found do not imply that Hwang-Noh theory is wrong. We must keep in mind that such theory was derived as a Newtonian limit and therefore we cannot expect it to work for neutron stars, which are relativistic systems. By the way, we can state that Hwang-Noh theory is ``less wrong" than the Newtonian one, when applied to the realm of neutron stars, since it at least predicts a stability curve.

On the other hand, the Neo-Newtonian theory had been proposed as an effective description of GR. Thus, it is natural that it works better than the Newtonian and Hwang-Noh theories when investigating the stability of a neutron star, despite its being less theoretically motivated.

Applying the four theories employed above for white dwarves, for which the Newtonian theory already works fine, we find basically the same result for the Chandrasekhar mass. This was expected since, for this kind of system, one has $p/\rho c^2 \sim 10^{-4}$. Indeed a very small correction.

\section{Discussion and conclusions}\label{sec:concl}

There are many attempts to modify the Newtonian equations in order to introduce typical relativistic effects. Some of them, in the spirit of the so-called Neo-Newtonian cosmology \cite{milne1, milne2, McCrea, Harrison, Lima:1996at, fabris}, look for reproducing typical relativistic effects, associated to the presence of pressure contributions at cosmological scales. While this proposal is very successful at the background cosmological level, it faces important limitations at the cosmological perturbative level (even at linear order) \cite{Lima:1996at, Reis:2003fs} and in the applications to compact stellar objects \cite{Fabris:2013psa, Fabris:2014ena, Oliveira:2014mka}. Such Neo-Newtonian formulations modify the usual Newtonian equations, in general, in an {\it ad hoc} way even if some thermodynamics and other special considerations can be evoked \cite{McCrea, Lima:1996at, fabris}.

Another approach is not to look for a Friedmann cosmology (or the TOV equations) reproduced fully at background and perturbative level by modifications of the Newtonian equations, but to try to extend the latter equations via introducing corrections, at a given perturbative order, due to full general relativistic equations, such that the Newtonian framework can be safely used in the regime of validity of the approximation. Such seems to be the proposal of Ref.~\cite{Hwang:2013sia}. In this approach, it can be expected that the resulting new ``Newtonian'' equations cannot be applied in principle to all scales in cosmology and for some compact objects in the relativistic regime (neutrons stars, for example). But, they can potentially be applied in very important problems, like the numerical simulations for structure formation, which demand the use of Newtonian physics and are implemented in the regime of cosmological small scales.

Here, we have revisited the proposal of Ref. \cite{Hwang:2013sia}. 
We have applied their equations in the background cosmological context, perturbative cosmological analysis and neutron stars configurations. In the first and third applications, serious limitations were found: in cosmology, the accelerated expansion occurs for positive pressure, an opposite effect to that found in GR; for neutron stars, the upper mass of these objects is more than one order of magnitude of that predicted by GR and suggested by observations \cite{obs}. However, it must be remarked that the instability region of the neutron stars are qualitatively reproduced.

On one hand, the predicted cosmological perturbative equations at small scales coincide with the full relativistic ones at linear order, see Ref.~\cite{Ma:1995ey}. This fact, opens the possibility to enlarge the use of the Newtonian framework in numerical simulations for the structure formation problem, using more reliable equations which reproduce relativistic effects even at perturbative level. However, it must be stressed that such simulations aim to study the formation of structure at deep non-linear level. In this sense, the analysis made here is just a first step in the program of constructing extended Newtonian equations that can be applied in such kind of problems.

{It is interesting to observe that both the investigated applications of background cosmology and stellar equilibrium are suggesting that the phenomenological Neo-Newtonian approach is physically more consistent with the full relativistic result than the Hwang-Noh approach. And this occurs despite the fact that the latter has been derived directly from GR, using a post-Newtonian approximation. On the other hand, the Hwang-Noh approach describes better than the Neo-Newtonian one the evolution of small fluctuations on small scales. For this reason, whenever using the full GR theory is not viable, the Neo-Newtonian theory (for large-scale cosmology or stellar equilibrium) and the Hwang-Noh theory (for small-scale cosmological perturbations) could help to get some hints on the problem at hand.}

\acknowledgments

The authors acknowledge CNPq and Fapes for support. They also thank L. Casarini, J.c. Hwang, H. Noh and W. Zimdahl for many enlightening discussions. {We are indebted with the anonymous referee for his/her enlightening comments which helped us to improve this paper.}

\bibliographystyle{JHEP}

\end{document}